\documentclass[aps,prl,twocolumn,superscriptaddress]{revtex4-1}
\usepackage{booktabs}
\usepackage{mathrsfs,amssymb,graphics,subfigure,threeparttable}
\usepackage{graphicx}
\usepackage{epstopdf}
\usepackage{amssymb}
\usepackage{enumerate}
\usepackage{amsmath}
\usepackage{fancyhdr}
\usepackage{bm}

\usepackage[squaren]{SIunits}
\usepackage{url}
\PassOptionsToPackage{hyphens,spaces,obeyspaces}{url}
\usepackage[hidelinks,breaklinks]{hyperref}

\begin{document}


\title{ Generating high quality ultra-relativistic electron beams using an evolving electron beam driver}



%
%
%
\author{T. N. Dalichaouch}
\affiliation{Department of Physics and Astronomy, University of California, Los Angeles, California 90095, USA}

\author{ X. L. Xu}
\affiliation{Department of Physics and Astronomy, University of California, Los Angeles, California 90095, USA}

\affiliation{SLAC National Accelerator Laboratory, Menlo Park, California 94025, USA}

\author{F. Li}
\affiliation{Department of Physics and Astronomy, University of California, Los Angeles, California 90095, USA}
\author{A. Tableman}
\affiliation{Department of Physics and Astronomy, University of California, Los Angeles, California 90095, USA}
\author{F. S. Tsung}
\affiliation{Department of Physics and Astronomy, University of California, Los Angeles, California 90095, USA}
\author{W. An}
\affiliation{Department of Physics and Astronomy, University of California, Los Angeles, California 90095, USA}
\author{W. B. Mori}
\affiliation{Department of Physics and Astronomy, University of California, Los Angeles, California 90095, USA}
\affiliation{Department of Engineering, University of California, Los Angeles, California 90095, USA}

%
%


\date{\today}

\begin{abstract}
A new method of controllable injection to generate high quality electron bunches in the nonlinear blowout regime driven by electron beams is proposed and demonstrated using particle-in-cell simulations.  Injection is facilitated by decreasing the wake phase velocity through varying the spot size of the drive beam and can be tuned through the Courant-Snyder (CS) parameters. Two regimes are examined. In the first, the spot size is focused according to the vacuum CS beta function, while in the second, it is focused by the plasma ion column. The effects of the driver intensity and vacuum CS parameters on the wake velocity and injected beam parameters are 
examined via theory and simulations. For plasma densities of  $\sim 10^{19}  ~\centi\meter^{-3}$, particle-in-cell (PIC) simulations demonstrate that peak normalized brightnesses $\gtrsim 10^{20}~\ampere/\meter^2/\rad^2$ can be obtained with projected energy spreads of $\lesssim 1\%$ within the middle section of the injected beam, and with normalized slice emittances as low as $\sim 10 ~\nano\meter$.    \end{abstract}


\maketitle



Over the past few decades, plasma-based acceleration (PBA), driven by either a laser pulse (LWFA) \cite{tajimadawson} or particle beam (PWFA) \cite{chendawson1985}, has attracted significant interest in compact particle accelerator and x-ray free-electron-laser (XFEL) applications due to the high accelerating fields $\sim10-100$ GV/m they generate \cite{hoganplasma4Gev10cm, blumenthal42GeV85cm, leemansplasma4GeV9cm, leemans20061GeV3cm, wang20132GeV7cm, hafz2008GeVbunches,litos2014high,adli2018acceleration,steinke2016multistage,PhysRevLett.122.084801}.  While the generation of ultra-relativistic electron beams through self-injection in an evolving plasma wake has been observed in LWFA experiments \cite{leemansplasma4GeV9cm, leemans20061GeV3cm, wang20132GeV7cm, hafz2008GeVbunches} and demonstrated in simulations  \cite{kalmykovevolvingplasmabubble2011, xu2005tightfocusedlaser,tsung2004selfinjection,lu2007lwfa1.5GeV}, the beams produced to date do not exhibit the sufficiently low energy spreads $\sigma_{\gamma}$ and high normalized brightnesses $B_n = 2I/\epsilon_n^2$ required to drive XFEL devices \cite{xfelbarletta2010} where $I$ and $\epsilon_n$ represent the current and normalized emittance, respectively. In recent years, electron injection schemes involving field ionization \cite{ ionizationinjconcept2006-2, ionizationinjconcept2006-3, xu2014colinearionization, li2013transversecollid, twocolorionization2014, 2012hiddingbeamlaserionization}
or the use of a plasma density down ramp (DDR) \cite{katsouleasdownramp1986,bulanovdownramp1998,sukdownramp2001, xu2017downrampinj,martinezdownramp2017} have shown tremendous potential for high quality beam generation for XFEL applications.

\begin{figure}[t]
\includegraphics[width=0.5\textwidth]{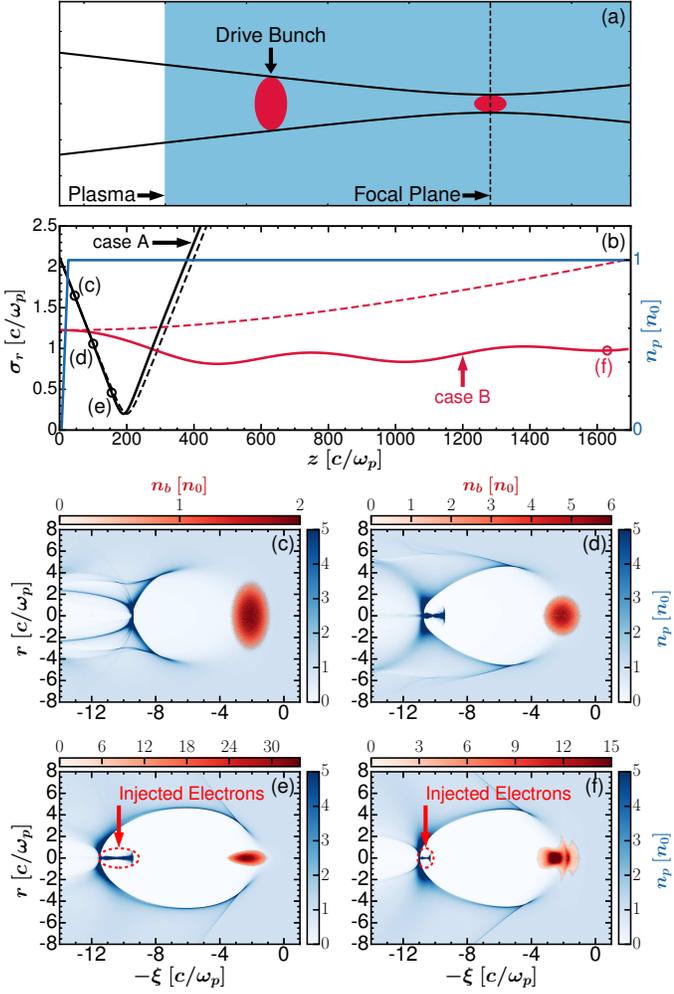}
\caption{\label{fig:schematic} (a) Schematic of self-injection in PWFA. Simulation results are shown in two regimes for for electron drivers with $\Lambda = 6$ focused into a plasma. (b) The projected spot size $\sigma_r$ of the driver is calculated from the simulation data (solid) and for vacuum propagation (dashed) for case A (black) and case B (red). The plasma density profile is shown in blue. The plasma wake is shown at propagation distances into the plasma of (c) $z= 45 \ c/  \omega_p$, (d) $z= 100 \ c/\omega_p$, (e) $z= 155 \ c/\omega_p$ for case A  and (f) $z= 1630 \ c/ \omega_p$ for case B.  }
\end{figure}

In this Letter, we propose and demonstrate a new method of controllable injection using an electron beam driver whose spot size is decreasing in the nonlinear blowout regime to control the wake phase velocity and hence induce electron trapping. As when using a DDR, this proposed method relies on gradually elongating the ion column or cavity length as the drive beam propagates. In this scheme, injection can be achieved by focusing the electron beam driver over spot sizes, $\sigma_r$, ranging from the scale length of the blowout radius $\sim r_b$ to spot sizes much less than $r_b$. A schematic of this process is shown in in Fig.~\ref{fig:schematic}(a). For spot sizes in this range, we will show that the ion column length and blowout radius slightly increase as $\sigma_r$ decreases and become insensitive to variations in $\sigma_r$ when $\sigma_r \ll r_b$. Therefore, self-injection can be induced by controlling the focusing of the drive beam.  This new approach is also a physically simpler alternative to injection methods such as DDR and Ionization, which require sharp density gradients or multiple drivers to produce high quality electron beams. Using OSIRIS simulations, we find for parameters expected to be achievable at FACET II that beams with $B_n \sim 10^{20}$ A/m\textsuperscript{2}/rad\textsuperscript{2}  could be produced.

The parameters that control this self-injection process are the peak normalized charge per unit length $\Lambda (\xi)  \equiv 4 \pi r_e \int_{0}^{r \gg \sigma_r} n_b (\xi) rdr$ where $\xi \equiv (ct-z)$ for a driver moving in the $\hat z$ direction,  and the Courant-Snyder (CS) parameters, $\beta$, $\alpha $, and $\gamma$ \cite{csparams}, where $\beta = \langle x^2 \rangle/\epsilon$, $\alpha = -\langle x x' \rangle/\epsilon$, $\gamma = -\langle {x'}^{2} \rangle/\epsilon$, and $\epsilon = \sqrt{\langle x^2 \rangle \langle {x'}^{2} \rangle -{\langle x x' \rangle}^2}$ is the transverse geometric emittance. The diffraction length for a particle beam in vacuum is $\beta^*\equiv \frac{\sigma_0^2}{\epsilon}$ where $\sigma_0$ is the focal spot size.

The proposed injection process relies on tuning how the spot size evolves. The initial evolution of the spot size, $\sqrt{\epsilon\beta}$,  can be estimated from Taylor expanding about its initial value as it enters the plasma,
\begin{align}
\label{eq:sigma}
\sigma_r = \sqrt{\epsilon \beta} = \sigma_i + \sigma_i' \Delta z + \frac{\sigma_i''}{2} \Delta z^2 ... 
\end{align}
where $\Delta z$ is the distance of the beam from the entrance of the plasma, $\sigma_i' = \sqrt{\epsilon_i \gamma_i} \frac{\alpha_i}{\sqrt{1+\alpha_i^2}}$ and $\sigma_i''$ for the part of the beam that resides in the fully formed wake can be obtained from $\sigma_r'' = \frac{\epsilon_n^2}{\gamma_b^2 \sigma_r^3} \left(1- \frac{\gamma_b^2 k_\beta^2 \sigma_r^4}{ \epsilon_n^2} \right)$ where $k_{\beta}\equiv \frac{k_p}{\sqrt{2\gamma_b}}$, $k_p \equiv \ \frac{\omega_p}{c}$, $\omega_p^2\equiv\frac{4\pi e^2 n_0}{m}$, $\epsilon_n \equiv \gamma_b \epsilon$ is the normalized emittance and $\gamma_b$ is used for the relativistic factor of the beam to differentiate it from the CS parameter, $\gamma$. Two distinct regimes exist depending on whether $\sigma_i'$ or $\sigma_i''$ terms dominate respectively.  These regimes can also be defined as when  $k_{\beta}\beta^*$  either $\ll$ or $\gg$  than unity.

These two regimes are illustrated in Fig.~\ref{fig:schematic} where results from  OSIRIS simulations are shown.  Simulations results presented in this Letter use the quasi-3D algorithm implemented in OSIRIS which expands the fields and currents into an arbitrary number of azimuthal modes m on an r-z grid \cite{lifshitz09,davidson16}. For the results presented here, m=0 and m=4 were used for symmetric and asymmetric drivers, respectively. We also use a new customized finite-difference solver to reduce numerical effects induced by the rapid rise in current during injection \cite{li2017, Xu2019}. The simulations used $ \Delta  z =  \Delta r = \frac{1}{128} \frac{c}{\omega_p}$, $\Delta t= \frac{1}{512}  \frac{1}{\omega_p}$ with 4 to 128 macro-particles per cell depending on the species and the number of azimuthal modes used.

In case A, a 10 GeV ($\gamma_b = 20000$) driver with $\Lambda=6$ is initialized with a centroid at $\xi = -2.1 ~c/\omega_p$, $\sigma_z = 0.7 \ c/\omega_p$, a normalized emittance $\epsilon_n \equiv \gamma_b \epsilon = 41.9 ~c / \omega_p $, and CS parameters  at the plasma entrance of $\beta_i \approx 2115 ~ c/\omega_p$ and $\alpha_i  \approx 10.5$. For these parameters $k_{\beta}\beta^* \approx.1$ and the evolution of the spot size is dominated by vacuum diffraction as seen by the agreement with spot size from the simulation with the vacuum evolution(dashed black) in Fig.~\ref{fig:schematic}(b). As shown in Fig. \ref{fig:schematic}(c), the wake is being excited in the nonlinear blowout regime when the beam enters the plasma. As the driver spot size decreases, in Fig. \ref{fig:schematic}(c)-(e) the bubble length also increases, thereby reducing the wake velocity $\gamma_{\phi}$ allowing the highest energy sheath electrons with $\gamma_z  \equiv (1-\beta_z^2)^{-1/2}> \gamma_{\phi}$ to be trapped at the rear of the wake. At spot sizes much less than the maximum blowout radius  $r_m \approx 2 \sqrt{\Lambda}c/\omega_p$ \cite{lu2006nonlinearphysplasma}, the wake wavelength and blowout radius approach their maximum values  in Figs. \ref{fig:schematic}(e) and further focusing does not produce any injection.

In case B, an electron drive bunch with the same length, centroid, and energy is focused to the entrance of the plasma with  $\epsilon_n \equiv \gamma_b  \epsilon = 24.5 ~c / \omega_p $, and CS parameters $\beta_i = \beta^* \approx 1225 ~ c/\omega_p$ and $\alpha_i  \approx 0$. For these parameters $k_{\beta}\beta^*=6.125$ and as seen in Fig \ref{fig:schematic}(b) the spot size evolution (solid red) is now dominated by the focusing force of the plasma ion column.  It therefore deviates significantly from the vacuum curve (dashed red). In this case the wake elongates as the beam focuses from the ion channel forces leading the self-injection as seen in Fig. \ref{fig:schematic}(f). This regime is more complicated because the ion channel is not fully formed along the entire driver and therefore only the rear of the beam is fully focused.  Each slice of the front of the beam oscillates at different betatron frequencies leading to the scalloping \cite{blumenthal42GeV85cm} seen in  Fig. \ref{fig:schematic}(f). Each scallop corresponds to a full betatron oscillation within the ion channel.

\begin{figure}[b]
\includegraphics[width=0.5\textwidth]{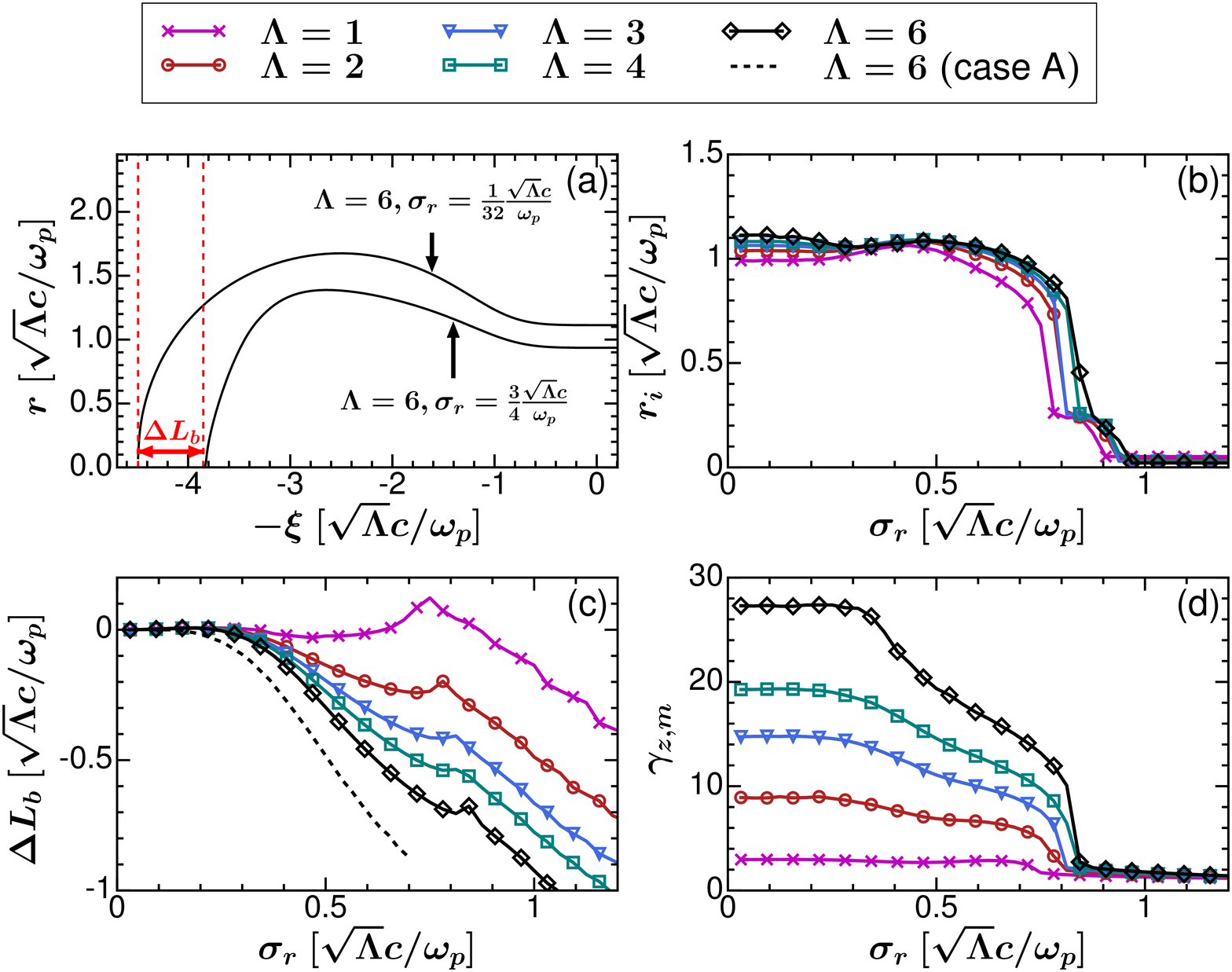}
\caption{\label{fig:sigma} (a) Trajectories of the most energetic sheath electrons for $\sigma_r = \frac{1}{32}\sqrt{\Lambda} c/\omega_p$ and  $\frac{3}{4}\sqrt{\Lambda} c/\omega_p$ when $\Lambda = 6$. The change in the bubble length $\Delta L_b$ is shown by the red dashed lines. (b) The initial position of the plasma sheath electrons $r_i$, (c) the change in bubble length $\Delta L_b$, and (d) the maximum sheath electron forward velocity $\gamma_{z,m}$ as a function of the spot size $\sigma_r$ for quasi-static plasma wakes. Lengths are normalized by $\sqrt{\Lambda} c/\omega_p$. Non-evolving electron drivers with energy $\gamma_b =20000$ and $\sigma_z =  0.7 \ c/\omega_p$ were used.  $\Delta L_b$ is also calculated from the evolving wake in case A [see Fig. \ref{fig:schematic}] and plotted in (c) with dashed black lines.
}
\end{figure}

In order to understand and quantify how the wake expansion depends on the evolution of the spot size, we performed numerous simulations using non-evolving drivers with varying $\sigma_r$ and $\Lambda$.  We tracked the trajectories of the sheath electrons to determine the blowout  radius $r_b$, the ion column length $L_b$, and the highest forward velocity $\gamma_{z,m}$ of sheath electrons at the rear of the wake. In Fig. \ref{fig:sigma}(a), we plot the trajectories of $r(\xi)$ for highest energy sheath electron for $\Lambda=6$ and two different driver spot sizes $k_p\sigma_r$=$\frac{1}{32}\sqrt{\Lambda}$ and $\frac{3}{4}\sqrt{\Lambda}$ respectively. These tracks clearly show that the ion channel size varies for these two spot sizes. In Figs.~\ref{fig:sigma}(b)-(d), for each simulation we also plot the initial  radius $r_i$ , deviation from the maximum bubble length $\Delta L_b(\sigma_r) \equiv L_b(\sigma_r) - L_b({\sigma_r \ll \sqrt{\Lambda} c/\omega_p})$, and maximum sheath electron forward velocity $\gamma_{z,m}\equiv (1-\beta_{z,m}^2)^{-1/2}$ for the most energetic electron for a variety of spot sizes and $\Lambda$.  The $\Delta L_b$ calculated from case A (dashed black) is also plotted in Fig. \ref{fig:sigma}(b) which is in reasonable agreement with the non-evolving results. The deviation can be explained from  beam loading effects from injected electrons. 

When $\sigma_r \gtrsim  \sqrt{\Lambda} c/\omega_p$, the driver produces a linear wake where the plasma electron perturbation is small, $\gamma_{z,m} \sim O(1)$, and the most energetic electrons originate near the axis $r_i \ll \sqrt{\Lambda} c/\omega_p$. The transverse force for a bi-Gaussian beam with a density profile $n_b \sim e^{-r^2/(2\sigma_r^2)}e^{-\xi^2/(2\sigma_z^2)}$  is $F_d = mc\omega_p(1-\beta_z) \frac{\Lambda(\xi,r)}{k_pr}$ where $\Lambda(\xi,r) \equiv \Lambda(\xi) \left[ 1-\exp{ \left( -\frac{r^2}{2\sigma_r^2} \right)} \right] $. In the large spot size limit, the transverse force of the driver on the sheath electrons is small since most of the driver charge resides outside the initial sheath position $\Lambda(\xi,r_i) \ll \Lambda(\xi)$ and the driver density remains below the particle crossing threshold as $n_b < 1.792 n_0$ \cite{lu2006nonlinearphysplasma}.

As the spot size approaches the particle crossing condition, typically around $\sim 0.75 - 0.85\ \sqrt{\Lambda} c/\omega_p$ depending on $\Lambda$, the wake response transitions from the linear regime to the blowout regime. This transition is signaled by a local maximum in $\Delta L_b$ and a rapid rise in $\gamma_{z,m}$ and $r_i$ due to formation of a thin plasma sheath. The most energetic sheath electrons now reside at initial positions $r_i \approx \sqrt{\Lambda} c/\omega_p$ [Fig. \ref{fig:sigma}(b)] which are insensitive to $\Lambda$.  As in the linear regime, the transverse force exerted on these electrons increases with the effective charge per unit length $\Lambda(\xi,r_i)$ as $\sigma_r$ decreases, creating larger blowout radii and bubble lengths as shown in Figs. \ref{fig:sigma}(a) and \ref{fig:sigma}(c). The sheath electrons also exhibit higher $\gamma_{z,m}$ [Fig. \ref{fig:sigma}(d)] as more background electrons are blown out by the driver, increasing the density spike and magnitude of $E_z$ at the rear of the wake. Although $\gamma_{z,m}$ increases significantly it is still orders of magnitude below $\gamma_b$ so trapping cannot occur. We also note that both $ \left|\frac{d(\Delta L_b)}{d\sigma_r}\right|$ and $\gamma_{z,m}$  increase with $\Lambda$ as seen in Fig. \ref{fig:sigma}(c)-(d). When $\sigma_r \ll  \sqrt{\Lambda} c/\omega_p$, the transverse driver force on the sheath electrons at $r_i$ peaks as $\Lambda(\xi,r_i) \rightarrow \Lambda(\xi)$, producing complete electron cavitation within the wake. In this limit, the bubble length $L_b$, blowout radius $r_m$, and $\gamma_{z,m}$ saturate at their maximum values and are insensitive to $k_p\sigma_r$.

For an evolving relativistic drive bunch $(\gamma_b \gg 1)$, the wake phase velocity is defined by
\begin{align}
\label{eq:velocity}
\beta_{\phi}  \approx 1 - \frac{dL_b}{dz}  = 1- \frac{dL_b}{d \sigma_r}\frac{d \sigma_r}{dz}. 
\end{align}
To calculate $\gamma_{\phi}$, we need to determine $\frac{dL_b}{d \sigma_r}$ and $\frac{d \sigma_r}{dz}$. We assume that $\frac{dL_b}{d \sigma_r} = \frac{d(\Delta L_b)}{d \sigma_r} $ depends only on the instantaneous spot size $\sigma_r$ and, therefore, can be calculated directly from Fig. \ref{fig:sigma}(c). The second term $\sigma_r^{\prime}$ is straightforward to obtain for case A. However, for case B $\sigma_r^{\prime}$ is more complicated to determine because the spot size evolution varies along $\xi$. The rear evolves as in a fully formed wake while the slices in the front have betatron frequencies that depend on the amount of blowout which leads to a scalloping behavior described above.

For simplicity, we analyze case A in which $\sigma_r$ evolves only as a function z and $\sigma_r (z)$ is well approximated by the vacuum CS parameters, $\sigma_r'=\frac{-\alpha\sqrt{\epsilon}}{\sqrt{\beta}}$. Using the beta function and the CS relation $\beta \gamma = 1+ \alpha^2$, we obtain an expression for the spot size evolution $\sigma_r^{\prime} = -\sqrt{\epsilon \gamma}     \frac{\alpha}{\sqrt{1+\alpha^2}} \approx -\sqrt{\epsilon \gamma_i} $, since $\gamma$ is constant in vacuum and $\alpha  \gg 1$ during most of the injection. 

\begin{figure}[t]
\includegraphics[width=0.5\textwidth]{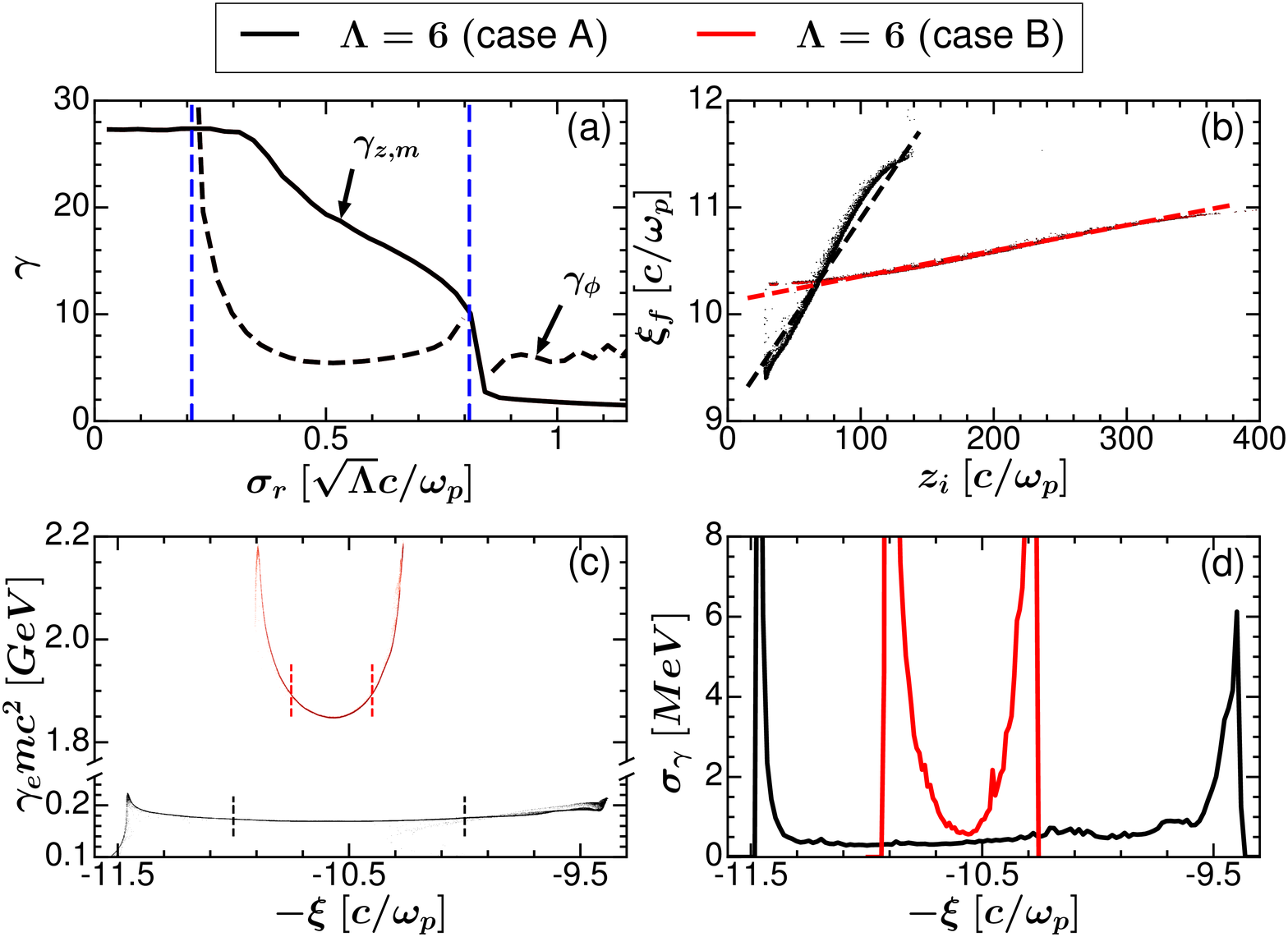}
\caption{\label{fig:injection} (a) $\gamma_{z,m}$ (solid) and $\gamma_{\phi}$ (dashed) calculated from Eq. (\ref{eq:velocity})  as a function of $\sigma_r$ for case A. The blue dashed lines enclose regions of injection where $\gamma_{z,m} > \gamma_{\phi}$. (b) The phase-space of injected electron charge in the $\xi_{f} - z_i$ plane for case A (black) and case B (red). The dashed lines are calculated from Eq. (\ref{eq:tracks}). (c) The phase space $\gamma - \xi$ and (d) slice energy spread $\sigma_{\gamma}$ of injected electrons at position $ z = 260 ~(1630) ~c/ \omega_p$ for case A (B). The injected beams are subdivided into 128 (64) slices for cases A (B), respectively, in (d).  } 
\end{figure}

We next approximate $\gamma_{\phi}\approx \sqrt{\frac{1}{2(1-\beta_{\phi})}}$ and use Eq. (\ref{eq:velocity}) to obtain,
\begin{align}
\label{eq:gamma_phi}
\gamma_{\phi}  \approx \sqrt{\frac{1}{2} \frac{1}{\frac{dL_b}{d \sigma_r}}  \frac{1}{\sigma_r'}}=  \sqrt{ \frac{\gamma_b}{-\frac{dL_b}{d \sigma_r}} \frac{\sigma_0}{2 \epsilon_n} } . 
\end{align}
From Fig. \ref{fig:sigma}(c) we can see that $\frac{dL_b}{d \sigma_r} \approx -1.76$ for $k_p\sigma_r=0.5\sqrt{\Lambda}$ and $\Lambda = 6$;  therefore, $\gamma_{\phi}$ ($\approx 5.2$) can be significantly reduced from $\gamma_b$ making injection possible. In Fig. \ref{fig:injection}(a), we plot $\gamma_{\phi}$ (dashed red) calculated from Eq. (\ref{eq:gamma_phi}) for case A against $\gamma_{z,m}$ (solid black) when $\Lambda = 6$ from Fig \ref{fig:sigma}(d). The vertical blue dashed lines enclose the region where the injection condition is satisfied $\gamma_{z,m} > \gamma_{\phi}$. As the driver focuses, self-injection begins near $\sigma_r \approx 0.81\sqrt{\Lambda}c/\omega_p$ as $\gamma_{z,m}$ rapidly increases and crosses $\gamma_{\phi}$. When $\sigma_r \approx 0.21 \sqrt{\Lambda}c/\omega_p$, the bubble length saturates $\frac{dL_b}{d\sigma_r} \rightarrow 0$ and injection terminates as $\gamma_{\phi}$ increases above $\gamma_{z,m}$. 

Since $L_b$ is a monotonically decreasing function of $\sigma_r$ in the blowout regime, there is a one-to-one mapping between the initial position $z_i$ and final position $\xi_f$ of injected electrons as seen in Fig. \ref{fig:injection}(b). This feature can lead to low absolute slice energy spreads as pointed out in  \cite{xu2017downrampinj}. The dependance of $\xi_f$ on $z_i$ is given by
\begin{align}
\label{eq:tracks}
 \frac{d\xi_f}{dz_i} \approx \frac{dL_b}{dz_i} \approx \kappa \frac{d\sigma_r}{dz_i}
\end{align}
where $\kappa$ is the average of $\frac{d(\Delta L_b)}{d\sigma_r} $ over the range $0.3-0.7\sqrt{\Lambda}c/\omega_p$ when $\Lambda = 6$  in Fig. \ref{fig:sigma}(c). Whereas  $\sigma_r^{\prime} \approx -\sqrt{\epsilon \gamma_i}$  is determined from the vacuum beta function in case A, an upper bound for $\sigma_r^{\prime}$ in case B can be estimated by integrating ${\sigma_r^{\prime\prime}} = \frac{\epsilon_n^2}{\gamma_b^2 \sigma_r^3} \left( 1-\frac{\gamma_b \sigma_r^4}{2 \epsilon_{n}^2} \right)$ and evaluating it at the matched condition $\sigma_r=\sigma_m = {\left( \frac{2 \epsilon_{n}^2 }{\gamma_b} \right)^{\frac{1}{4}} }$.  To account for axial variations in the betatron frequency in case B, we also assume a simple model in which half of the drive bunch undergoes betatron oscillations while the remaining half is approximately non-evolving during injection. In Fig. \ref{fig:injection}(b), excellent agreement is observed between the injected electron phase space $(\xi_f,z_i)$ and Eq. (\ref{eq:tracks}) in both cases. After some optimal acceleration distances, flat phase space distributions  in $(\gamma_e ,\xi)$ are achieved over some regions of the injected beams as shown in Fig. \ref{fig:injection}(c). Within the regions enclosed by dashed lines in Fig. \ref{fig:injection}(c), average energies of 169.7 MeV and 1.86 GeV and projected energy spreads of 1.1\% and 0.7\% for cases A and B, respectively. In both cases, slice energy spreads as low as $\sim$ 0.5 MeV are also observed along significant regions of the injected electron beams in Fig. \ref{fig:injection}(d).


\begin{figure}[b]
\includegraphics[width=0.5\textwidth]{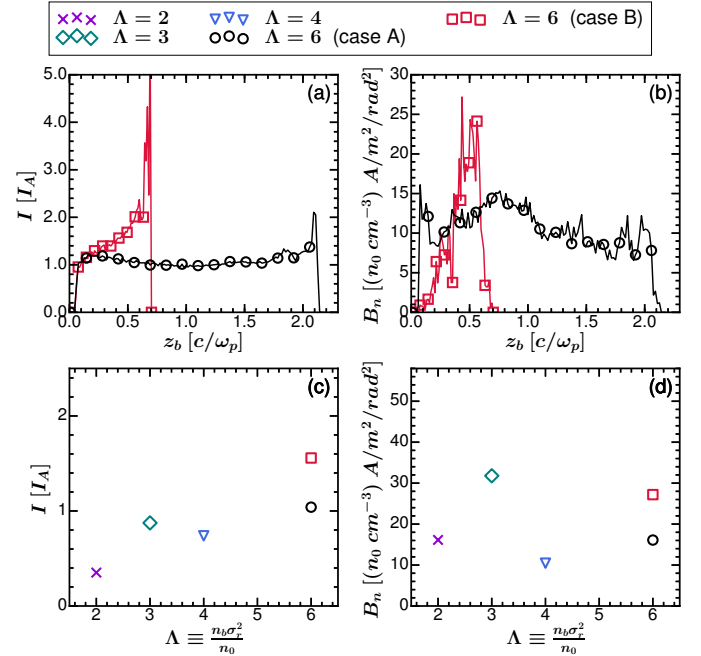}
\caption{\label{fig:beam_focus_200} The beam slice parameters of the injected electrons for case A and case B:  (a) the current, in units of the Alfven current $I_A$ and (b) the normalized brightness $B_n$ in units of $ [(n_0 ~\centi\meter^{-3})  ~\ampere/\meter^2/\rad^2]$. The injected beams are subdivided into 128 (64) slices for cases A (B), respectively. (c) Slice current corresponding to the (d) peak normalized brightness of the injected electron beam as a function of the driver parameter $\Lambda$. Each marker corresponds to a unique simulation. Approximately 95 \% of injected electrons are used when calculating the beam parameters.}
\end{figure}

The slice currents and normalized brightnesses of the final injected beams are calculated and plotted in Fig. \ref{fig:beam_focus_200}(a)-(b). While the injected beam duration in case A is nearly three times that of case B, larger slice currents of $I \gtrsim 17-34 ~\text{kA}$ are observed across most of the injected beam in case B. The large variations in the current profile at the front of the injected beam in case B can be attributed to a more slowly evolving wake and beam loading effects from the initial electron injection as noted in \cite{xu2017downrampinj}. In Fig. \ref{fig:beam_focus_200}(b), peak normalized brightnesses of $\sim 16 \ \text{and} \ \sim 27 \ [n_0~\centi\meter^{-3}] ~ \ampere/\meter^2/\rad^2$ are found for cases A and B, respectively.  In Figs. \ref{fig:beam_focus_200}(c)-(d), peak slice brightnesses and corresponding slice currents are plotted for various simulations ranging from $\Lambda$ of 2 to 6. In the results shown, peak brightnesses of at least $\gtrsim 10^{21} ~\ampere/\meter^2/\rad^2$ are achieved across all values of $\Lambda$ for plasma densities of $n_0 \sim 10^{20}  ~\centi\meter^{-3}$ with slice currents ranging from $ 6-26 ~\text{kA}$.

At FACET II, it is anticipated that drive beams with currents in the range of 50-150 kA and durations of $\sim$ 3 fs (1$~\micro \meter$) will be available \cite{facet2016}.  These parameters match the simulations presented here for $n_0 \sim 10^{19}  ~\centi\meter^{-3}$, indicating that peak brightnesses as high as $ 10^{20} ~\ampere/\meter^2/\rad^2$ and normalized emittances as low as $\sim 10~\nano\meter$ could be produced. We have also carried out simulations with spot size asymmetries ($1-\sigma_x /\sigma_y$) of as high as 15$\%$ and the results indicate self-injection still occurs but with a reduction in brightness of less than an order of magnitude up to the maximum asymmetry. 

This work was supported by US NSF grant Nos. 1500630 and 1806046, US DOE grant No. DE-SC0010064, and FNAL sub award  544405, and by DE NSFC Grants No. 11425521, No. 11535006, No. 11375006, and No. 11475101, and Thousand Young Talents Program. The simulations were performed on Blue Waters,  the National Energy Research Scientific Computing Center (NERSC), and Hoffman2 at UCLA. The corresponding authors are X. L. Xu (xuxinlu04@gmail.com) and T. N. Dalichaouch (tdalichaouch@gmail.com).

\bibliographystyle{apsrev4-1}

\bibliography{refs_thamine}
\end{document}